\documentclass[12pt]{article}
\usepackage{latexsym}
\usepackage{amsmath}
\usepackage[dvips]{graphicx}
\usepackage{textcomp}

\newcommand{\be}{\begin{equation}}
\newcommand{\ee}{\end{equation}}
\newcommand{\ba}{\begin{array}}
\newcommand{\ea}{\end{array}}

\author{Fabio Cardone $^{1,2,3}$, Roberto Mignani $^{3}$,
Andrea Petrucci $^{3,4,5*}$\\ \\ $^{1}$Istituto per lo Studio dei
Materiali Nanostrutturati (ISMN — CNR) \\ Via dei Taurini - 00185
Roma, Italy
\\ $^{2}$GNFM, Istituto Nazionale di Alta Matematica "F.Severi" \\
Citt\`a Universitaria, P.le A.Moro 2 - 00185 Roma, Italy \\
$^{3}$Dipartimento di Fisica "E.Amaldi", Universit\`a degli Studi
"Roma Tre" \\ Via della Vasca Navale, 84 - 00146 Roma, Italy \\
$^{4}$LNF-Istituto Nazionale di Fisica Nucleare
\\ Via Enrico Fermi 40 - 00044 Frascati (Rome), Italy
\\ * Correspondent author: petruccia@fis.uniroma3.it}
\date{}
\title{Experimental and phenomenological comparison between
Piezonuclear reactions and Condensed Matter Nuclear Science
phenomenology.}
\begin{document}
\maketitle \abstract{The purpose of this paper is to place side by
side the experimental results of Piezonuclear reactions, which have
been recently unveiled, and those collected during the last twenty
years of experiments on low energy nuclear reactions (LENR). We will
briefly report the results of our campaign of piezonuclear reactions
experiments where ultrasounds and cavitation were applied to
solutions of stable elements. These outcomes will be shown to be
compatible with the results and evidences obtained from low energy
nuclear reaction experiments. Some theoretical concepts and ideas,
on which our experiments are grounded, will be sketched and it will
be shown that, in order to trigger our measured effects, it exists
an energy threshold, that has to be overcome, and a maximum interval
of time for this energy to be released to the nuclear system.
Eventually, a research hypothesis will be put forward about the
chance to raise the level of analogy from the mere comparison of
results up to the phenomenological level. Here, among the various
evidences collected in LENR experiments, we will search for hints
about the overcome of the energy threshold and about the mechanism
that releases the loaded energy in a suitable interval of time.}

\section{Evidences of Piezonuclear reactions}
All of the experiments that we have carried out, involved the
application of ultrasounds with a frequency of 20 kHz and suitable
power (most of the times around 100 W) to liquids. These liquids
were either bidistilled deionised water~\cite{carmig4,carmig5,
carmig6} or solutions of bidistilled-deionised water and some
standard chemical elememts~\cite{carmig7,card}. In the two
experiments reported in ~\cite{carmig4,carmig5} samples of
bidistilled and deionised water were subjected to
cavitation\footnote{It is necessary to warn the reader that
ultrasounds and cavitation do not have to be read as the mean to
produce hot fusion confined in a collapsing bubble as it was for
other experiments conducted by other teams. In the third paragraph
the differences between our scope and their will be clarified} for
different intervals of time: 210 minute long cavitation induced an
increase of the proportion of a few high mass number stable isotopes
(including uranium); four successive cavitations, each one lasting
for 10 minutes, with 15 minutes between a cavitation and the next
one, gave rise to an increase of the proportion of a few nuclear
species within the particular atomic mass range 238$<$M$<$264
(including transuranic elements). In the third
experiment~\cite{carmig6}, aimed at detecting elements in the so-
called rare-earth mass range, ICP mass spectrometry of the solution
was performed during five successive cavitations, each one lasting
for 15 minutes (with 15-minutes intervals between a cavitation and
the next one). The ICP-MS analysis gave evidence of a significant
peak corresponding to a nuclide with atomic mass (137.93$\pm$0.01)
amu and half-life 12$\pm$1 sec, identifed as $Eu^{138}_{63} $. It is
known that the abundance on Earth of stable Eu is less than 1.06 ppm
(the natural Eu is a mixture of two isotopes, $Eu^{151}_{63}$ with a
percentage abundance of 47.77\% and $Eu^{153}_{63}$ with a
percentage abundance of 52.23\%). The candidate identied during the
third experiment, $Eu^{138}_{63}$, does not exist in nature; it is
an articial radionuclide (discovered in 1995-97~\cite{euro}) that
can be produced at the present time in nuclear reactors and by
synchrotrons. There are two ways whereby $Eu^{138}_{63}$ can be
produced: by nuclear fission or by nuclear fusion. The former
process requires less energy. However, from the results of the first
two experiments, the quantity of heavy nuclei, which can produce
$Eu^{138}_{63}$ by nuclear fission, is much smaller (by two-three
orders of magnitude) than that of the intermediate nuclei that can
produce it by fusion. It turns out that fusion intermediate of
elements is more likely and besides it is the only possible
explanation of the changes in concentration of stable elements,
induced by cavitation, observed in the first experiments
\footnote{The words fission and fusion have been used here. However,
it is becoming evident that the outcomes of these piezonuclear
reactions experiments, along with those involving low energy nuclear
reactions, have nothing to do with the established definitions of
fission and fusion. In this sense, these two words have to be
interpreted as disgragation of nuclei and union of nuclei
respectively. In the following other words will be used like
nucleolysis and nucleosynthesis}. The ionising radiation
measurements, by LR115 detectors, that were carried out during the
application of ultrasounds in all of these experiments, did not
provide any evidence of ionising radiation above the background
level. Further experiments were designed in order to try and detect
possible neutron emission. We subjected to cavitation, bidistilled
deionised water, solutions of Lithium, Aluminium, and Iron. No
evidences of neutrons were collected with water Lithium and
Aluminium. Iron, conversely, did produce neutrons every time and the
mutually exclusive experiments stressed that, being equal all the
experimental conditions, it was just the presence of Iron to bring
about emission of neutrons. Quite bewilderingly, such a heavy and
stable element\footnote{Iron is at the top of the curve binding
energy per nucleon.}, whose Coulomb barrier is huge, was the trigger
of unusual neutron emission\footnote{Neutrons were detected by
different techniques: bubble detectors, CR39 with Boron layer, Boron
Trifluoride. The evidences collected were compared with neutron
signals from standard sources (AmBe) and fast neutron nuclear
reactor}. A further unusual circumstance was the lack of the gamma
radiation that usually comes along with the emission of neutrons. A
further experiment was performed in order to verify the effects of
these new mechanisms induced by ultrasounds and cavitation on
radioactive nuclei~\cite{torio,torcom1,torcom2}. The evidences
indicated that the initial quantity of Thorium became half in a
interval of time 10000 times faster than Thorium half life. However
it turned out that this process was not a mere acceleration of the
usual Thorium decay by emission of alpha particles, since the number
of tracks on the CR39 detectors that monitored the radioactive
process was not compatible with this possibility.

\section{Evidences of Low Energy Nuclear reactions}

More than 20 years have gone since the first announcement of Cold
Fusion by Martin Fleischmann and Stanley Pons. Since then, despite
the aversion that this subject received, an incredible amount of
experiments have been carried out in order to reproduce the
announced cold fusion effects. Different techniques have been tested
and great improvements in reproducibility have been achieved. We
will try and group the apparently anomalous results obtained during
these years in order to point out the possible analogies among these
outcomes and those collected in the experiments of piezonuclear
reactions. Since the goal of this paper is only to show a new
possible perspective and promote discussion on it, comprehensiveness
is not our main target. In order to summarize the results we will
refer to the book "The Science of Low Energy Nuclear Reactions" by
Edmund Storms~\cite{storms}. Independently from the the method used
to induce LENR (Electrolyte, Plasma, Laser, Diffusion, Fuse,
Ambient, Bombard, Biological) it becomes clear that in all of
experiments there were clear signs of transmutations and that most
of the times the resulting products were Iron, Zinc Copper, Nickel,
Chromium and other nuclides with comparable mass and binding energy
per nucleon (none of them can be considered a light element).
Besides, the nuclides belonging to the substrates used in all of the
experiments had considerable atomic mass ranging from 48 amu for Ti
or 58 amu for Ni up to Pd with 106 amu and further on to W with 184
amu and Au with 197 amu. As to the environment in which the
substrate was immersed, it contained different substances and
chemical compounds which contained much lighter nuclides like H, D,
Li, Na, K, C, N, O, Cl, and sometimes other heavier ones compatible
with the atomic mass of the nuclides of the substrate. Both in
piezonuclear reactions experiments and in LENR experiments there are
transmutations that involve heavy mass number nuclides and  produce
other intermediate and heavy mass number nuclei\footnote{In
piezonuclear reactions the heavy nuclides which can be placed side
by side to the LENR substrate are those contained in the alloy of
the sonotrode}. These transmutations took place neither by fusion
(Coulomb barriers would be too high) nor by fission since the
evidences did not show any signs typical of these reactions, like
neutron emission or prompt gamma rays or presence of easily
detectable radionuclides\footnote{Of course, being neutrons very
rare in both of the type of reactions transmutations by neutron
capture are ruled out as well or at least very unlikely.}. In LENR
neutron emission was very low and infrequent while in piezonuclear
reactions it was not infrequent but nevertheless it was low and
certainly not compatible with knonw nuclear reactions, first of all
because of the lack of prompt gamma rays and second because no gamma
rays from neutron capture by hydrogen was detected either. As to
other kinds of radiation emitted during LENR, many different types
were detected, which, however have not help in identifying clear
common features among the different experiments and techniques.
Among all of them it is worth noting that some teams detected a
strange radiation showing unknown features and
behaviour~\cite{storms,uru1} which, from our point of view, could be
put beside the strange lack of gamma rays which, at least from
hydrogen neutron capture, should be emitted. As it will become clear
from the next section, it was not our goal to perform extra power or
heat measurements during piezonuclear reaction experiments, and
hence no comparison can be made on this ground.

\section{Local Lorentz Invariance Breakdown}

The theoretical background, on which our experiments have been
designed and carried out, is based on the concept of breakdown of
Local Lorentz Invariance (LLI)~\cite{carmig1,carmig3}. LLI is a
symmetry of the laws of Physics which locally, i.e. in a
sufficiently small region of space-time, have to stick to the
framework of Special Relativity~\cite{will}. This statement has some
interwoven consequences on the mathematical form of the laws and on
the structure of local space-time. Our theory concentrates on the
structure of local (microscopical) space-time (flat and rigid
according to LLI) when LLI is broken. The theory has been built from
a phenomenological basis, in the sense that the coefficients of the
local Minkowski metric tensor are hypothesized to be function of the
energy of the process. The form of these functions have been
determined by analysing through this formalism possible anomalous
outcomes of experiments probing different fundamental interactions.
The dependence of the metric tensor on the energy of the physical
process means that local space-time is certainly no more rigid and
moreover that its geometry can be deformed, just like a blanket
which can be furrowed and creased by the energy of a hand. The main
consequence of this locally deformed space-time is that it takes an
active part in the dynamics of the physical process whose features
and flow deeply depart from their usual appearance. The theory
predicts that the space-time of strong interactions begins to be
deformed when the energy of the process overcomes a threshold of
energy equal to 367.5 GeV~\cite{carmig1,carmig3}\footnote{This value
seems quite big, but it is necessary to refer it to a microscopical
region (the active reagion or NAE) and to the macroscopical amounts
of energy pumped in the every type of LENR experimental setup}.
Besides it clearly states that there is no isochrony between the
time of the experimenter and that of the hadronic process. To put it
in a more practical way, this means that in order to deform
space-time around a nucleus (or nuclei) and hence trigger
"anomalous" processes, in the fashion of those presented above from
piezonuclear reactions, one has to find a microscopical mechanism
that loads an amount of energy higher than the energy threshold and
then it is capable to release it in an suitable interval of time or
in other words it is capable of a suitable power (energy divided by
time). This theoretical background shows that our experiments was
not in wake of the LENR ones, but their target was to obtain some
evidences that would corroborate the two predictions mentioned above
about the threshold energy and the release mechanism\footnote{This
explains why we have never used deuterated substance, never look for
Helium or Tritium and never attempted to measure the presence of
extra heat.}.

\section{LLI and anomalous nuclear processes}

It has been shown that the results of LENR experiments and those of
piezonuclear reactions are compatible. Thus, despite the apparent
diversity of the experimental setups and conditions, it is possible
to hypothesize that similar outcomes are brought about by similar
microscopical mechanisms that trigger alike anomalous nuclear
processes. Now, let us evidence the phenomenological aspects that,
within piezonuclear experimental setups, fulfill the two conditions
mentioned above. Once that these aspects are clear we will try and
analyze some LENR setups and evidences in order to make out similar
features.

First of all it is important to state that piezonuclear reactions
have nothing to do with sonofusion despite the common starting
point, i.e. ultrasounds and cavitation. The two research tracks
diverge from the theoretical, phenomenological and experimental
points of view~\cite{piez_sono}\footnote{Even a short explanation of
the differences between the two research tracks would be a
digression from the main target of this paper. We only mention here
that the collapse of the bubble is not a mean to increase the
temperature of its content, but, conversely, a mean to accelerate
the ions trapped on its surface towards each other. For a concise
but clear explanation please refer to the web site
http://www.newnuclear.splinder.com/tag/towards+clean+nuclear+energy}.
The collapse of a generic bubble under the huge pressure of
ultrasounds, that induces shock-waves on its surface, is regarded as
the microscopical mechanism to fulfil the two requirements. Let us
recall the evidences obtained in the experiments where neutron
emission has been detected~\cite{carmig7,card}. We subjected to
cavitation bidistilled deionised water, a solution of Lithium
Chloride, a solution of Aluminium Chloride, a solution of Iron
Chloride and a solution of Iron Nitrate~\cite{carmig7}. Only the two
solutions containing Iron emitted neutrons, apparently, without
gamma rays above the background level. Besides, the emission of
neutrons did not begin as soon as the ultrasounds were turned on.
The first neutron evidences began to appear after 40 - 50 minutes
since ultrasounds had been turned on. These two facts, Iron response
and the existence of a delayed emission of neutrons\footnote{delayed
emission refers to the 40 - 50 minutes mentioned above and has
nothing to do with radioactive decay of radionuclides} (which has
been a constant evidence in every experiment) can be referred to the
predicted energy threshold. It is possible to hypothesize that the
40 - 50 minutes were the time to reach and overcome the energy
threshold. This hypothesis has to be completed with an important
detail. Overcoming the energy threshold needs to be referred to the
context and the environment where the process takes place and in
particular to the types of nuclides involved. The experimental
conditions were the same for the three types of nuclides, Lithium,
Aluminium and Iron. Only the solutions with Iron emitted bursts of
neutrons. This can be interpreted in terms of the two conditions
mentioned above (about the overcoming of the energy threshold and
the interval of time to release it) by saying that Iron, for some
reason, that can be now only conjectured, possesses those peculiar
features that make it fulfill, within the experimental conditions,
the two prescriptions. The conjecture is that, the collapse of the
bubble concentrates energy in a smaller and smaller region of space
(which is actually spacetime), making the energy density higher and
higher. In this region of spacetime nuclear species are forced. The
overcoming of the threshold is achieved by the complementary
contributions of the external energy (ultrasounds) and internal
energy, i.e. that of the nuclides taking part in the collapse. The
first preliminary clue is that the higher the atomic mass the less
external energy and the shorter time interval it takes to deform
locally the spacetime. This must be considered only a rough and
highly incomplete picture\footnote{Actually, it not clear whether
one of the variables is atomic mass or binding energy or binding
energy per nucleon.}. More variables are certainly involved. It
seems that the quantity of bosonic and fermionic isotopes for each
nuclide could play an important role as well\footnote{About the last
conjecture, the picture is a lot more complicated since one should
distinguish between nuclides that transmute and the results of
transmutations}. Despite the incomplete picture, it is possible to
state that the results of the experiments on piezonuclear reactions
(transmutations and neutrons without gamma rays) described in the
first part of this paper are brought about by pressure combined with
a mechanism that allows an abrupt release of energy within a
suitable interval of time. The pressure is produced by ultrasounds,
the mechanism is the collapse of the bubble\footnote{Critical
parameters are the dimension of the bubble and the number of ions of
the specific nuclide present on the its surface}. Their synergy
along with nuclides with high atomic mass form the Nuclear Active
Environment (NAE) in our experiments.\\ Once that
pressure\footnote{piezo comes from the greek word piezein which
means to press}  and bubble collapse have been identified as the
conditions to generate a NAE in piezonuclear reactions, let us try
and see if it is possible to spot within different experimental
setups, where ultrasounds are not used, the counterparts of pressure
and bubble collapse. Many different setups have been used in LENR
experiments and evidences of nuclear activity have been collected in
almost all of them. Due to the initial character of the clues that
this paper puts forward, it certainly does not want to be
exhaustive. In this sense, it is interesting to concentrate on the
evidences collected by an electrochemical cell with Pd/D
co-deposition which, first of all, is the LENR technique more
similar to that used by Fleischmann and Pons and more over it seems
the farthest, in terms of experimental conditions, from the
piezonuclear reactions setup. No pressure seems to be involved, no
ultrasounds, no cavitations, and from a mere visual and audible
perspective, it seems much quieter. All of the evidences reported
below refer to Pd/D co-deposition experiments and the corresponding
control tests. In particular, we refer to the experimental setup and
procedure used by Mosier-Boss et al. which is described
in~\cite{bosssetup}. In their experiments the metals used for the
cathode were Ni, Ag, Pt, Au and the solution contained $PdCl_2$ and
 $LiCl_2$ or $KCl_2$ in heavy water. In control experiments
$PdCl_2$ was substituted for $CuCl_2$. The procedure comprised two
phases. During the first part of the electrolysis, Palladium is
reduced on the cathode (Ni, Ag, Pt, Au) and gets plated onto it
together with Deuterium. Once the Palladium is plated out, the
second phase begins. During this part an external static electric or
magnetic field can be either applied or not to the cell and the
cathodic current is increased. The details that we are reporting
here can be extracted from
~\cite{bosssetup,bosstrans,bosstriple,bosscharged} and the
references cited in them. Our purpose is to try and look through
them in order to make out those features that may corroborate the
hypothesis of the existence in the electrolytic technique of a
mechanism that may fulfill the two requirements mentioned above
about the energy threshold and releasing time. Bubble collapse is a
far from equilibrium process where energy is locally concentrated
around the surface of the bubble and suddenly released with great
intensity. In this regard, we will have to look at the whole
electrolytic technique and search for leads pointing toward these
three characteristics (far from equilibrium, local loaded
energy,local abrupt release of it). In~\cite{bosstrans} the
researchers state that "the experimental protocol covers three time
periods". The first is the co-deposition of Pd/D on the cathode
which takes place for several hours at different increasing
currents; the second period is called "stabilization of the system",
when, at increasing different currents, for several hours, the Pd/D
ratio of the co-deposition is let distribute uniformly on the
cathode; the third period of time is used "to put the system in a
far from equilibrium condition". This is done by applying an intense
static electric or magnetic field and by letting the electrolysis
proceed by keeping increasing the current from time to time. It is
fairly reasonable to consider that a co-deposition of Pd/D is in
itself an unstable structure in which atomic bonding between
Palladium atoms are deformed, stretched, and weakened by the
presence of Deuterium. To confirm this, one can put forward the fact
that LENR experiments have begun to be carried out in the last few
years not by loading Deuterium in bulky Palladium, but rather by
co-depositions Pd/D or by nanostructured Pd. These methods produce
greater quantities of anomalous evidences than those with solid
Pd~\cite{storms,bosssetup,celani}. Let us look at this unstable
structure of Pd/D from the point of view of the hypothesized NAE in
which heavy mass number nuclides play a fundamental role. From this
perspective, Palladium would be the main reactant (precursor) while
the role of Deuterium, loaded among Palladium atoms, would be to
load mechanical energy within Pd\footnote{This perspective must be
considered only a conjecture sustained only by some experimental
evidences of piezonuclear reactions in solids, which will be
presented soon. Of course one might wave the evidence that
macroscopic outcomes, in terms of transmutations, ionizing
radiation, extra heat, were only be obtained where Deuterium was
used and that, viceversa, when Hydrogen was used the results were
much less evident. However, Deuterium cold fusion explains only very
hardly the anomalous outcomes of LENR and CMNS experiments.
Conversely, always remaining of a conjectural ground, it is possible
to hypothesize that, once the microscopical spacetime has been
deformed thanks to presence of a heavy nucleus, like Pd, lighter
nuclides may fall into the deformation and take part in the non
Minkowskian nuclear reaction. In this non Minkowskian sense
Deuterium contributes with one neutron more than Hydrogen.}. Anyway,
leaving to further experiments the conjecture of Palladium as the
main reactant, that the loading of mechanical energy in Pd is
brought about by Deuterium, is confirmed by experiments in which the
solution contained $CuCl_2$ instead $PdCl_2$. Copper does not
absorbs Deuterium, no lattice deformation is brought about, no
mechanical energy gets loaded and hence no reactions take
place~\cite{bosssetup}. The statements by Storms~\cite{storms} in
his book seem to move in the same direction. He reports that "...the
basic material used as cathode is not active initially even when it
is made of Palladium - activation is required. Nevertheless, the
base material does affect the morphology and subsequent activity of
the deposited layer...". The co-deposition or the nano-structures
together with heavy atomic mass nuclides, used in the substrates
(Pd, Au,Ag,Pt,Ni...), are some of the ingredients to generate a NAE
and in particular those ingredients that allow to store an amount of
energy higher than the threshold mentioned above. All of these
conjectures can be considered a sound lead\footnote{A lead in the
sense that deeper investigation, experimental, theoretical and
phenomenological will have to be carried on from this point.} from
the experimental conditions about the possible fulfilment of the
first requirement of LLI breakdown hypothesis: existence of an
energy threshold. We concentrate now on the second requirement: the
mechanism to release this loaded energy. Very inspiring in this
regard, is the discussion in the papers~\cite{bosstrans,bossdeform}.
In~\cite{bosstrans} the discussion is about the shape change that
was noticed in some areas of the cathode at the end of the
electrolysis. The unstable structure at the cathode after Pd/D
co-deposition with energy loaded mentioned above is a very
complicated configuration of electric layers which, if the current
were stopped, would be in a reasonable macroscopical equilibrium but
certainly a feeble microscopical one. The application of the intense
electric or magnetic field and the increase of the current shake the
microscopical equilibrium and produce hot spots randomly distributed
in position and in time all over the cathode. As stated by the
researchers in the paper~\cite{bosstrans}, "...a conductor placed in
an electric field cannot remain in a stable equilibrium and a
negative pressure acts on its surface...". This pressure is the
consequence of forces applied to the surface of the conductor,
which, due to the Gauss' theorem, is affected by them in its bulk as
well. The feeble microscopical equilibrium is due to the complex
electrical relation existing among the three components of the
cathodic system: the substrate (Ni, Ag, Pt, Au), the co-deposition
Pd/D and the solution. First of all this dynamically unstable
relation is established by the electric potential differences which
can be synoptically represented by the Fermi level and the redox
potential of the three components and second by the presence of
Deuterium and its continuous supply. Deuterium moves within
Palladium and this exposes the latter to local gradient of pressure.
Along with it, the flow of an increasing current alters the electric
equilibrium and generate local forces and hence pressures. Besides,
one might think that if the electrolytic current is not stable but
presents brief and maybe small variations (small spikes), they
produce gradients of magnetic field which, on their side, induce
gradients of pressure\footnote{Although one may think that these
gradients of magnetic field are small, they have to be imagined
applied to microscopical regions}.\\We reckon that, from the picture
presented above, it is fairly reasonable to hypothesize that within
the electrolytic technique with Pd/D co-deposition there exist local
experimental conditions compatible with those found in piezonuclear
reaction experiments where cavitation took place. As a bubble is
local frail inhomogeneity within a liquid which can be squeezed and
deflated\footnote{We remind that in our phenomenological model the
generic bubble deflates while being squeezed by a shockwave. In
other words it is not treated as mean to compress the gas contained
in it in order to reach hot fusion conditions.} by squeezing it by
ultrasounds, the locally frail structure of Pd/D co-deposition
presents local inhomogeneities or hollows (e.g. gradients of the
density of Pd and/or D atoms) whose sudden and violent\footnote{The
rapidity and force of the collapse will be two of the many
parameters that will have to be estimates, however the violence here
refers to the microscopical region where pressure is applied}
collapse can be induced by bringing the systems to a far from
equilibrium condition (electric or magnetic field, periodical
increase of current, continuous flow of Deuterium with the
co-deposited layer.) As Iron atoms (not light elements) are
entrapped in the interface gas/liquid of bubbles and are launched
against each other during the collapse and forced in a smaller and
smaller volume of spacetime until the energy threshold is reached
and overcome in a precise interval of time, Palladium atoms might
endure similar processes as hollows in Pd/D co-deposition are made
collapse. One more mental picture can be added to the scenario.
Let's imagine that these conditions are established in one point of
the Pd/D co-deposition, and that these processes take place
there\footnote{There is no need for imagination, at least for the
processes, since this is what comes out from all of the evidences.}.
These processes in one point might trigger other similar processes
in points of Pd/D not too far from it. It would be a bit as if a
balanced seesaw were outweighed on one side (external cause like
current or fields) and, as it touches the ground, it triggered a
slight explosion, which would thrust this part upwards and the
opposite one downwards which, in turn, would trigger a second
explosion, and so on.\\ Many an evidence has been collected by
different experimental techniques compatible with processes or NAE
that are randomly distributed all over Palladium (hot spots,
concentrated evidences of charged particles, concentrated presence
of anomalous transmutation products). Control experiments with
Copper instead of Palladium are in favor of these hypotheses since
Copper does not absorb Deuterium and hence no suitable structures
with hollows or gradients of density are formed~\cite{bosssetup}.
Besides, when bulk Palladium is used instead of Pd/D co-deposition,
some evidences are obtained but far less than with the
latter~\cite{bosssetup}. In~\cite{storms} cracks are indicated as
"...the only environments obviously common to all successful
experiments...". Always in~\cite{storms} it is said that Palladium
expands as it is loaded with Deuterium and cracks of different
dimensions form during this process. Cracks are said to be present
in Pd/D co-depositions or Palladium black. Some questions are raised
as to how cracks are involved in the cold fusion process, or as to
how cracks operate to allow Coulomb barrier penetration, or as to
how dimensions of cracks influences the formation of a NAE.
Eventually, cracks are said to be good candidates to be a NAE.
According to our theory and hence according to our hypotheses, the
NAEs that form in our experiments and in all of the LENR/CMNS
experiments do not trigger well known nuclear processes and we
believe that this is a fact more than corroborated by loads of
anomalous evidences. Moreover, our theory predicts, although at an
initial level, that these new nuclear processes cannot be account
for by the well known laws of (nuclear) physics. Conversely, since
local Lorentz invariance breakdown and spacetime deformation take
place, all the concepts like Coulomb barrier, that belong to a flat
spacetime, may not be suitable to describe both qualitatively and
quantitatively these new processes. Nuclear spacetime deformation
which, according to us is the heart of these phenomena, has been
discovered to be brought about by pressure and an energy releasing
mechanism (cavitation in our experiments) which succeed in creating
those conditions mentioned above that are the overcoming of the
367.5 GeV energy threshold and the release of this energy in a
precise interval of time. Under this point of view, cracks are the
equivalent of bubbles. However, the important part is how rapidly
and forcibly these cracks (hollows) are made
collapse~\cite{carpinstrain,carpinarxiv} and in this sense the
dimensions of the crack play their role. Before concluding, we would
like to point out one more thing. We said that, high atomic mass
nuclides contribute to facilitate the overcoming of the energy
threshold when they are forced in a smaller and smaller region of
spacetime. This qualitative picture may induce to think that
starting from Palladium, one should obtain heavier nuclei like in a
sort of new type nuclear fusion. If it were like this, the evidences
would be against. In~\cite{bosstrans} it is reported of
transmutations whose products were Aluminium, Magnesium, Chlorine,
Silicon which are all lighter nuclides than Palladium. Since these
transmutations are thought to be brought about by spacetime
deformation, it is possible that a heavy nuclide be ripped apart
into lighter nuclides by tidal forces i.e., in more picturesque way,
as an astronaut would be as he were falling into a blackhole.

\end{document}